\documentclass[sigconf]{acmart}

\usepackage{multirow}
\usepackage{soul}
\usepackage{balance}

\newtheorem{theorem}{Theorem}[section]
\newtheorem{lemma}[theorem]{Lemma}

\AtBeginDocument{%
  \providecommand\BibTeX{{%
    \normalfont B\kern-0.5em{\scshape i\kern-0.25em b}\kern-0.8em\TeX}}}

\copyrightyear{2023}
\acmYear{2023}
\setcopyright{acmlicensed}\acmConference[SIGIR '23]{Proceedings of the 46th International ACM SIGIR Conference on Research and Development in Information Retrieval}{July 23--27, 2023}{Taipei, Taiwan}
\acmBooktitle{Proceedings of the 46th International ACM SIGIR Conference on Research and Development in Information Retrieval (SIGIR '23), July 23--27, 2023, Taipei, Taiwan}
\acmPrice{15.00}
\acmDOI{10.1145/3539618.3591649}
\acmISBN{978-1-4503-9408-6/23/07}

\begin{document}

\title{Collaborative Residual Metric Learning}

\author{Tianjun Wei}
\email{tjwei2-c@my.cityu.edu.hk}
\orcid{0000-0001-7311-7101}
\affiliation{%
  \institution{City University of Hong Kong}
  \city{Kowloon}
  \country{Hong Kong}
}
\author{Jianghong Ma}
\authornote{Corresponding author.}
\email{majianghong@hit.edu.cn}
\orcid{0000-0002-0524-3584}
\affiliation{%
 \institution{Harbin Institute of Technology}
 \city{Shenzhen}
 \country{China}}

\author{Tommy W. S. Chow}
\email{eetchow@cityu.edu.hk}
\orcid{0000-0001-7051-0434}
\affiliation{%
  \institution{City University of Hong Kong}
  \city{Kowloon}
  \country{Hong Kong}
}

\renewcommand{\shortauthors}{Wei and Chow, et al.}

\begin{abstract}
    In collaborative filtering, distance metric learning has been applied to matrix factorization techniques with promising results. However, matrix factorization lacks the ability of capturing collaborative information, which has been remarked by recent works and improved by interpreting user interactions as signals. This paper aims to find out how metric learning connect to these signal-based models. By adopting a generalized distance metric, we discovered that in signal-based models, it is easier to estimate the residual of distances, which refers to the difference between the distances from a user to a target item and another item, rather than estimating the distances themselves. Further analysis also uncovers a link between the normalization strength of interaction signals and the novelty of recommendation, which has been overlooked by existing studies. Based on the above findings, we propose a novel model to learn a generalized distance user-item distance metric to capture user preference in interaction signals by modeling the residuals of distance. The proposed CoRML model is then further improved in training efficiency by a newly introduced approximated ranking weight. Extensive experiments conducted on 4 public datasets demonstrate the superior performance of CoRML compared to the state-of-the-art baselines in collaborative filtering, along with high efficiency and the ability of providing novelty-promoted recommendations, shedding new light on the study of metric learning-based recommender systems.
\end{abstract}





\maketitle

\section{Introduction}
A growing interest in Collaborative Filtering (CF) \cite{Adomavicius2005} has been seen in both academia and industry. The main challenge of CF is to handle the interactions between users and recommended targets, often named as items \cite{Deshpande2004}. Since this interaction can be naturally represented by an interaction matrix, factorization-based models have become one of the common paradigms in CF. The most basic factorization-based model is the traditional Matrix Factorization (MF) \cite{Koren2009, Rendle2009}, where user-item preferences are computed by lightly designed dot product of user and item embeddings. Against its simplicity in design, MF is considered to lack the ability to capture higher-order user-item relationships \cite{Wang2019}, which is questioned and improved by emerging Graph Convolutional Network (GCN) models recently \cite{Wang2019, He2020, Wei2023}. In contrast, signal-based models treat the interaction matrix as signals for each user \cite{Shen2021}, and learn relationships between items. A straightforward signal-based approach is the linear autoencoder \cite{Ning2011, Steck2019}, which models item-item relations as a square matrix of linear mappings, achieving competitive performance against factorization-based models with high training efficiency. A recent study \cite{Shen2021} adopts graph signal processing to handle user features and proposes a signal-based graph filtering framework, also yielding competitive performance.

In recent years, growing attention has been paid to recommender systems based on metric learning \cite{Xing2002, Weinberger2009}.  By learning a distance metric, metric learning drives the distance between samples to comply with their similarity or dissimilarity. In this way, metric learning has a natural fit with CF, which aims to explore the relationship between users and their interacted and uninteracted items. The emergence of many CF models based on metric learning \cite{Hsieh2017, Park2018, Li2020, Wu2020, Bao2022} is consistent with this intuition, where the most representative one is the Collaborative Metric Learning (CML) \cite{Hsieh2017}. In \cite{Hsieh2017}, the authors point out that the dot product in the traditional factorization-based models violates a crucial rule in a valid distance metric, the \textbf{triangle inequality}, and therefore fails in capturing fine-grained user-item relationship information. To overcome the deficient, CML is proposed as a new framework for estimating user-item preferences via Euclidean distances between embedding vectors rather than dot product. CML establishes a connection between factorization-based model and metric learning in Euclidean space, and its superior performance compared to MF models inspires a promising direction. However, metric learning on Euclidean space lacks the ability to accommodate signal-based models, where the user features are expressed using a fixed interaction signals. To our advantage, research on metric learning is not limited to the Euclidean distance, but has been extended to the generalized Mahalanobis distance \cite{Ghojogh2022}. The paradigm of metric learning in a generalized scenario is similar to the signal-based CF models,  thereby tempting us to explore their connections. To this end, we are eager to investigate the following research questions:
\begin{itemize}
    \item With the definition of generalized Mahalanobis distance, can a signal-based model learn a valid distance metric? If so, what conditions need to be satisfied?
    \item If a signal-based model can learn a distance metric, how the characteristics of that metric will affect CF task in terms of performance and other metrics, like novelty?
\end{itemize}

To answer the raised questions, we carry out an analysis on existing signal-based models. First, on the basis of the ranking-based feature in CF tasks, we conduct investigation on the relative relationships of the distance between users and different items, rather than the values of the absolute distance. Such differences, also called as the \textbf{residual} \footnote{In this paper, "residual" refers to the difference in distances or scores between the user's pair with the target item and the pair of other items.} of distance, is shown to be associated with the item-item relationship in the signal-based models. Specifically, when the symmetry and zero diagonal conditions of item-item relationship matrix are satisfied, the residual of generalized Mahalanobis distances between different user-item pairs are explicitly related to the residuals of the preference scores produced by a signal-based model. With this observation, we are able to learn a signal-based model to take advantage of metric learning and capture fine-grained user-item relationships. Besides, we further explore the role of the normalization strength of the interaction signals in signal-based models and demonstrate its importance in mitigating popularity bias of recommender systems and promoting the novelty of the recommendations, which is overlooked by existing studies. This motivated us to propose a new signal-based model to learn a generalized distance metric, aiming to derive novelty-promoting recommendations.

Based on the above analysis, we finally propose a novel model for CF task, named Collaborative Residual Metric Learning (CoRML). Specifically, by adopting widely used triplet margin loss in metric learning,  we propose a simplified residual margin loss to maximize the residual of preference score between interacted and uninteracted items. Through this loss, CoRML can learn a generalized Mahalanobis distance metric under any normalization strength, which is an extension of existing signal-based models. By tuning the normalized strength of the interaction signal, CoRML is able to generate highly accurate recommendations while ensuring the novelty of the recommendations. Then, by converting the original pairwise learning objective to point-wise, and approximating the dynamically updated ranking of items though a novel proposed ranking weight, CoRML is further improved in training efficiency compared with existing metric learning models in CF. Extensive experiments on 4 public datasets shows that CoRML is able to produce novelty-promoting recommendations with ensuring superior performance when comparing with state-of-the-art baselines. The PyTorch implementation code of the proposed CoRML model is publicly available at \url{https://github.com/Joinn99/CoRML}.

To summarize, the contributions of this paper are listed below:
\begin{itemize}
    \item We reveal the connection between existing signal-based models and metric learning, and identify critical factors in such models for promoting the novelty of recommendations.
    \item To address the limitations of existing models, we propose a novel CoRML model that efficiently models the residuals of the distance to capture user preferences.
    \item Extensive experiments demonstrate the superiority of CoRML over state-of-the-art CF models in terms of recommendation performance, training speed, and novelty.
\end{itemize}

\section{Preliminaries}
\subsection{Problem Formulation}
In this paper, we focus on the Collaborative Filtering (CF) task with user-item implicit feedbacks. Suppose the user set $\mathcal{U}$ and item set $\mathcal{I}$. For each user $u \in \mathcal{U}$, the non-empty set $\mathcal{I}_u \subseteq \mathcal{I}$ denotes the items that user $u$ has interacted with. Then, given the interacted item set $\mathcal{I}_u$, the goal of CF is to generate an $K$-item candidate set $\mathcal{I}^{'}_u (K) \subset \mathcal{I} \setminus \mathcal{I}_u$ as the recommended items for user $u$.

\subsection{Metric Learning}
Given a collection of data points $\{\mathbf{x}_i\}_{i=1}^N \subset \mathbb{R}^d$ with size $N$, metric learning aims to learn a distance metric to decrease the distance between similar points and increase the distance between dissimilar points \cite{Xing2002}. The similarity of data points is typically determined by the priori information, such as the class labels in a classification problem. In metric learning, a widely adopted distance metric is generalized Mahalanobis distance \cite{Ghojogh2022}. The generalized Mahalanobis distance between data points $\mathbf{x}_i$ and $\mathbf{x}_j$ is defined as
\begin{equation}
d(\mathbf{x}_i, \mathbf{x}_j)=\sqrt{(\mathbf{x}_i- \mathbf{x}_j)^T\mathbf{W}(\mathbf{x}_i- \mathbf{x}_j)},
\end{equation}
where $\mathbf{W}$ is a symmetric positive semi-definite (PSD) weight matrix to ensure the learned distance metric is valid and do not violate the triangle inequality:
\begin{equation}
d(\mathbf{x}_i, \mathbf{x}_j) \leq d(\mathbf{x}_i, \mathbf{x}_k) + d(\mathbf{x}_k, \mathbf{x}_j).
\end{equation}
When $\mathbf{W}$ is the identity matrix $\mathbf{I}$, the distance measured is the Euclidean distance between $\mathbf{x}_i$ and $\mathbf{x}_j$. To derive $\mathbf{W}$, metric learning generally formulates an optimization problem by measuring the similarity and dissimilarity of sample points. One of the classic solutions is to minimize the distances between similar data points and maximize the distances between dissimilar data points \cite{Ghodsi2007}:
\begin{equation}
\begin{aligned}
\underset{\mathbf{W}}{\mathrm{minimize}}&{\:\: \sum\limits_{(\mathbf{x}_i, \mathbf{x}_j) \in \mathcal{S}}d(\mathbf{x}_i, \mathbf{x}_j) -\sum\limits_{(\mathbf{x}_i, \mathbf{x}_j) \in \mathcal{D}}d(\mathbf{x}_i, \mathbf{x}_j)}\\
s.t.&\:\: \mathbf{W} \succeq 0, tr(\mathbf{W})=\alpha,
\end{aligned}
\end{equation}
where $\mathcal{S}$ and $\mathcal{D}$ denote the similar and dissimilar pairs of data points, respectively.  The trace of $\mathbf{W}$ is restricted to be a positive constant $\alpha$ to prevent the trivial result $\mathbf{W}=\mathbf{0}$. To date, numerous studies have extended the field of metric learning, while the core ideas mentioned above are still retained \cite{Ghojogh2022}.

\subsection{Collaborative Metric Learning}
\label{SubSec:CML}
With the growing interest in research on recommender systems, various studies have focused on the role of metric learning in CF tasks. To capture the user preference towards different target items, CF models draw their attention to deal with the historical user-item interactions. Given user set $\mathcal{U}$ and item set $\mathcal{I}$, the historical user-item interactions with implicit feedbacks can be represented as an interaction matrix $\mathbf{R}\in\{0,1\}^{\lvert \mathcal{U} \lvert \times \lvert \mathcal{I} \lvert}$, which is defined as
\begin{equation}
R_{ui}= 
\begin{cases}
    1, & \text{if interaction }(u, i)\text{ is observed,}\\
    0, & \text{otherwise.}
\end{cases}
\end{equation}
In CF, the classical matrix factorization (MF) \cite{Koren2009} techniques have been widely adopted to deal with $\mathbf{R}$. MF factorizes $\mathbf{R}$ to generate $d$-dimension embedding vectors $\mathbf{E}_U \in \mathbb{R}^{\lvert \mathcal{U} \rvert \times d}$ for users and $\mathbf{E}_I \in \mathbb{R}^{\lvert \mathcal{I} \rvert \times d}$ for items by solving the following optimization problem:
\begin{equation}
\underset{\mathbf{\mathbf{E}_U, \mathbf{E}_I}}{\mathrm{minimize}}{\:\frac{1}{2}\lVert \mathbf{R}-\mathbf{E}_U\mathbf{E}_I^T \rVert_F^2 + \frac{\theta}{2}(\lVert \mathbf{E}_U\rVert_F^2 + \lVert \mathbf{E}_I\rVert_F^2)},
\end{equation}
where $\theta$ is a hyperparameter for regularization. Learning embedding vectors for users and items is followed by a great deal of research and became a mainstream setting in CF tasks \cite{He2020, Mao2021}.

Next, we will introduce how metric learning is used to refine CF models. Collaborative Metric Learning (CML) \cite{Hsieh2017} is first proposed to formulate user preferences towards items from the perspective of distance metrics. CML considers the Euclidean distance between the user embedding vector $\mathbf{e}_u$ and the item embedding vector $\mathbf{e}_i$ as the score of the user's preference to the item, which is defined as:
\begin{equation}
d(\mathbf{e}_u, \mathbf{e}_i)=\lVert \mathbf{e}_u - \mathbf{e}_i \rVert_2=\sqrt{\mathbf{e}_u^T\mathbf{e}_u+\mathbf{e}_i^T\mathbf{e}_i-2\mathbf{e}_u^T\mathbf{e}_i}.
\label{Eq:CML_Distance}
\end{equation}
In Eq. \eqref{Eq:CML_Distance}, the term $\mathbf{e}_u^T\mathbf{e}_i$ is the score function in MF models. Therefore, CML essentially adds the embedding norm to the preference score to avoid violating triangle inequality, which is proved to be effective in retaining fine-grained preference information \cite{Hsieh2017}. In CML, the interaction matrix $\mathbf{R}$ is used to group similar and dissimilar pairs of users and items. All user and item pairs $(u, i)$ that have $R_{ui}=1$ are considered similar pairs, whose distances are optimized to be smaller than other pairs through triplet margin loss:
\begin{equation}
\underset{\mathbf{\mathbf{E}_U, \mathbf{E}_I}}{\mathrm{minimize}}{\:\sum \limits_{(u,i)\in \mathcal{S}}\sum \limits_{(u,j)\in \mathcal{D}}(d^2(\mathbf{e}_u, \mathbf{e}_i)-d^2(\mathbf{e}_u, \mathbf{e}_j)+\zeta)_+},
\label{Eq:Margin_Loss}
\end{equation}
where $\zeta$ is the hyperparameter denotes the margin of distance, $\mathcal{S}$ and $\mathcal{D}$ denote the set of pairs of user and interacted item, the set of pairs of user and uninteracted item, respectively. In general, triplet margin loss aims to pull all interacted items closer to user, and push the uninteracted items farther away to a safety margin \cite{Hsieh2017}.

Although CML and subsequent studies \cite{Hsieh2017, Li2020, Bao2022} are conducted based on the distance metric in the Euclidean space, they can still be translated into generalized Mahalanobis distance by the transformation of the feature space. Suppose the feature space of users and items is an identity matrix $\mathbf{I} \in \mathbb{R}^{(\lvert \mathcal{U} \rvert + \lvert \mathcal{I} \rvert) \times (\lvert \mathcal{U} \rvert + \lvert \mathcal{I} \rvert)}$, the distance between user $u$ and item $i$ can be equivalently represented as
\begin{equation}
d(\mathbf{i}_u, \mathbf{i}_i)=\sqrt{(\mathbf{i}_u- \mathbf{i}_i)^T\mathbf{E}\mathbf{E}^T(\mathbf{i}_u-\mathbf{i}_i)},
\label{Eq:CML_GM}
\end{equation}
where $\mathbf{E} \in \mathbb{R}^{(\lvert \mathcal{U} \rvert + \lvert \mathcal{I} \rvert) \times d}$ is the concatenated embedding vectors of users and items, and the weight matrix $\mathbf{W}=\mathbf{E}\mathbf{E}^T$ is a rank-$d$ symmetric PSD matrix.

\subsection{Signal-based Models}
In Section \ref{SubSec:CML}, we show that the traditional CML models based on MF can be interpreted as a special case of generalized Mahalanobis distance metric learning. Like Eq. \eqref{Eq:CML_GM}, the features of both users and items are represented as an identity matrix, while the parameters are the low-rank approximation of the weight matrix $\mathbf{W}$. Here, it is apparent that CML does not take full advantage of the interaction data to build the feature space. The interaction matrix $\mathbf{R}$ is only used to classify similar and dissimilar data points, leaving the feature space to be orthogonal identity matrix. 

The same weakness also exists in traditional MF models, which is questioned and improved by considering the interaction matrix as signals of users \cite{Chen2021}. Here, we consider the row-wise signals in $\mathbf{R}$, representing the interacted items for each user. Suppose the feature space of users and items are $\mathbf{P}\in \mathbb{R}^{\lvert \mathcal{I} \rvert \times \lvert \mathcal{U} \rvert }$ and $\mathbf{Q}\in \mathbb{R}^{\lvert \mathcal{I} \rvert \times \lvert \mathcal{I} \rvert }$ defined as follows:
\begin{equation}
\mathbf{P}=\mathbf{D}_I^{-t}\mathbf{R}^T, \: \mathbf{Q}=\mathbf{D}_I^{t},
\end{equation}
where $\mathbf{D}_I=diag(\mathbf{1}^T\mathbf{R})$ is the degree matrix of items, $t$ is a factor for the normalization strength of signal. Suppose $u$-th column of $\mathbf{P}$ is $\mathbf{p}_u$ and $i$-th column of $\mathbf{Q}$ is $\mathbf{q}_i$, signal-based models learn a weight matrix $\mathbf{C}\in \mathbb{R}^{\lvert \mathcal{I} \rvert \times \lvert \mathcal{I} \rvert}$ and generate preference score by
\begin{equation}
y_{ui}=\mathbf{p}_u^T\mathbf{C}\mathbf{q}_i.
\label{Eq:Graph_Signal_Score}
\end{equation}
One of the most well-known signal-based recommendation approaches is the \textbf{linear autoencoder} \cite{Ning2011, Steck2019, Steck2020}. In general, linear autoencoder learns the weight matrix by solving a constrained optimization problem \cite{Steck2019}:
\begin{equation}
\begin{aligned}
\underset{\mathbf{C}}{\mathrm{minimize}}&{\:\:\frac{1}{2}\lVert \mathbf{R} - \mathbf{R} \mathbf{C} \rVert_F^2},\:\: s.t.\: diag(\mathbf{C})=\mathbf{0},
\end{aligned}
\label{Eq:LAE}
\end{equation}
where diagonal zero constraint is used to prevent trivial solution $\mathbf{C}=\mathbf{I}$. On the other hand, a recent study formulates the CF problem as a low-pass graph filtering process \cite{Chen2021} and propose a \textbf{graph filtering model}. It first performs singular value decomposition (SVD) on the normalized interaction matrix $\mathbf{D}_U^{-\frac{1}{2}}\mathbf{R}\mathbf{D}_I^{-\frac{1}{2}}$, which is also known as graph Laplacian matrix. Here $\mathbf{D}_U =diag(\mathbf{R}\mathbf{1})$ is the degree matrix of users. Then, the right singular vectors $\mathbf{V}\in \mathbb{R}^{\lvert \mathcal{I} \rvert \times k}$ with top-$k$ largest singular values are applied to approximate $\mathbf{C}$, and generate recommendations by
\begin{equation}
\mathbf{R}\mathbf{D}_I^{-\frac{1}{2}}\mathbf{VV}^T\mathbf{D}_I^{\frac{1}{2}}.
\label{Eq:GF}
\end{equation}
Taken together, both linear autoencoders and graph filtering model can be considered as a special case of signal-based models, with different formulations of $\mathbf{C}$ and different settings in normalization strength $t$, respectively. 

\section{Identifying the Relationship Between Signal-based models and Metric Learning}
In reviewing the discussion of metric learning, signal-based models share many characteristics with Mahalanobis distance metric. This makes us curious if the user and item relationships learned in signal-based model can be expressed as generalized Mahalanobis distances. If so, what conditions do the weight matrix need to satisfy? These questions will be discussed in the following pages.
\subsection{Can Signal-based Models Learn a Distance Metric?}
\label{SubSec:CanLearn}
Suppose the feature spaces of users and items are $\mathbf{P}$ and $\mathbf{Q}$ respectively, then the generalized Mahalanobis distance of user $u$ and item $i$ can be represented as
\begin{equation}
\begin{aligned}
d(\mathbf{p}_u, \mathbf{q}_i)=&\:\sqrt{(\mathbf{p}_u-\mathbf{q}_i)^T\mathbf{W}(\mathbf{p}_u-\mathbf{q}_i)}\\
=&\: \sqrt{\mathbf{p}_u^T\mathbf{W}\mathbf{p}_u+\mathbf{q}_i^T\mathbf{W}\mathbf{q}_i-2\mathbf{p}_u^T\mathbf{W}\mathbf{q}_i}.
\label{Eq:Graph_Signal_Distance}
\end{aligned}
\end{equation}
In Eq. \eqref{Eq:Graph_Signal_Distance}, each terms are shared in different pairs of distances. For example, the term $\mathbf{p}_u^T\mathbf{W}\mathbf{p}_u$ exists in all distances that involve user $u$. In metric learning, what we are primarily concerned with is the relative relationship of distances between similar and dissimilar nodes. This is consistent with the objective of the CF task, which achieves recommendations by sorting the preference scores for users and different items. Therefore, we next focus on the residual of distance between different pairs of user and item. Suppose that user $u$ and two items $i$ and $j$, the residual of squared distances $d^2(\mathbf{p}_u, \mathbf{q}_i)$ and $d^2(\mathbf{p}_u, \mathbf{q}_j)$ is derived as
\begin{equation}
\begin{aligned}
\Delta d^2 =&\:d^2(\mathbf{p}_u, \mathbf{q}_i) - d^2(\mathbf{p}_u, \mathbf{q}_j)\\
=&\:\mathbf{q}_i^T\mathbf{W}\mathbf{q}_i-\mathbf{q}_j^T\mathbf{W}\mathbf{q}_j-2\mathbf{p}_u^T\mathbf{W}(\mathbf{q}_i-\mathbf{q}_j)\\
=&\:W_{ii}(d_i^{2t}-2R_{ui})-W_{jj}(d_j^{2t}-2R_{uj})-2\mathbf{p}_u^T\mathbf{H}(\mathbf{q}_i-\mathbf{q}_j),
\label{Eq:Diff_Dist}
\end{aligned}
\end{equation}
where $d_i$ is the degree of item node $i$, also refers to $i$-th element of $\mathbf{D}_I$, $\mathbf{H}$ is a diagonal-zero matrix contains the non-diagonal values of $\mathbf{W}$, also known as the \textit{Hollow matrix}. To learn a valid generalized Mahalanobis distance metric, $\mathbf{W}$ has to be symmetric PSD. The next proposition shows that the above condition is easy to satisfy with only ensuring the symmetry of $\mathbf{H}$.
\begin{theorem}
For any $n \times n$ symmetric hollow matrix $\mathbf{H}$ and positive vector $\mathbf{x} \in \mathbb{R}_+^n$, there always exists a positive value $\omega$ such that $\mathbf{H}+\omega diag(\mathbf{x}) \succeq 0$.
\end{theorem}
\begin{proof}
Let $\omega=max_{1\leq i \leq n} \frac{h_i}{x_{i}}$, where $h_i$ is the sum of absolute value of the non-diagonal entries in the $i$-th row of $\mathbf{H}$. With Lemma \ref{Lemma:Gershgorin}, we can show that every eigenvalue of $\mathbf{W}=\mathbf{H}+\omega diag(\mathbf{x})$ lies within at least one of the range $[\omega x_i-h_i, \omega x_i+h_i]$. As $\omega x_i\geq h_i$ for all $1 \leq i \leq n$, we have
\begin{equation}
  \lambda_k(\mathbf{W}) \geq 0, 1\leq k \leq n,
\end{equation}
where $\lambda_k(\mathbf{W})$ is the $k$-th eigenvalue of $\mathbf{W}$. Hence, $\mathbf{W} \succeq 0$ is proved.
\end{proof}
\begin{lemma}[Gershgorin circle theorem for symmetric real matrix]
Let $\mathbf{A}$ be a symmetric square real matrix, $r_i$ is the sum of the absolute values of the non-diagonal entries in the $i$-th row of $\mathbf{A}$:
\begin{equation}
  r_i = \sum \limits_{j \neq i} \lvert A_{ij} \rvert.
\end{equation}
Then every eigenvalue of $\mathbf{A}$ lies within at least one of the range $[A_{ii}-r_i, A_{ii}+r_i]$.
\label{Lemma:Gershgorin}
\end{lemma}
Then, let $diag(\mathbf{x})=\mathbf{D}_I^{-2t}$, the residual of distance in Eq. \eqref{Eq:Diff_Dist} is derived as
\begin{equation}
    \frac{1}{2}\Delta d^2=\begin{cases}
    -\:\tilde{y}_{uij}, &i \notin \mathcal{I}_u, j \notin \mathcal{I}_u\\
    -\:\tilde{y}_{uij}-\omega d_i^{-2t} , & i \in \mathcal{I}_u, j \notin \mathcal{I}_u \\
    -\:\tilde{y}_{uij}-\omega(d_i^{-2t}-d_j^{-2t}), & i \in \mathcal{I}_u, j \in \mathcal{I}_u,
\end{cases}
\label{Eq:Diff_GSM}
\end{equation}
where $\tilde{y}_{uij}=y_{ui}-y_{uj}=\mathbf{p}_u^T\mathbf{H}(\mathbf{q}_i-\mathbf{q}_j)$. Here we name $\tilde{y}_{uij}$ as \textbf{preference residual}, as $y_{ui}$ can be produced by signal-based models like Eq. \eqref{Eq:Graph_Signal_Score} as the preference score. Then, from Eq. \eqref{Eq:Diff_GSM}, there are several findings:
\begin{enumerate}
    \item When a user has not interacted with both item $i$ and $j$, $\Delta d^2$ can be obtained by $\tilde{y}_{uij}$ without error.
    \item When a user has only interacted with the item $i$, there is always a positive margin between $-2\tilde{y}_{uij}$ and $\Delta d^2$ .
    \item When a user has interacted with item $i$ and $j$, the difference between $\Delta d^2$ and $-2\tilde{y}_{uij}$ is varied based on $d_i$ and $d_j$, which is a constant $0$ only when $t=0$.
\end{enumerate}
And conclusions can be drawn from these findings:
\begin{itemize}
    \item \textbf{Conclusion 1}: The ranking of the generalized Mahalanobis distances between user and all uninteracted items can be accurately derived by the preference residual. This is essential for CF tasks, which typically generates recommendations by sorting the preference scores of uninteracted items.
    \item \textbf{Conclusion 2}: When the user's interacted items are considered, the preference residual is in general biased compared to the residual of generalized Mahalanobis distance.
\end{itemize}

\begin{figure}[t!]
  \centering
  \includegraphics[width=3.0in]{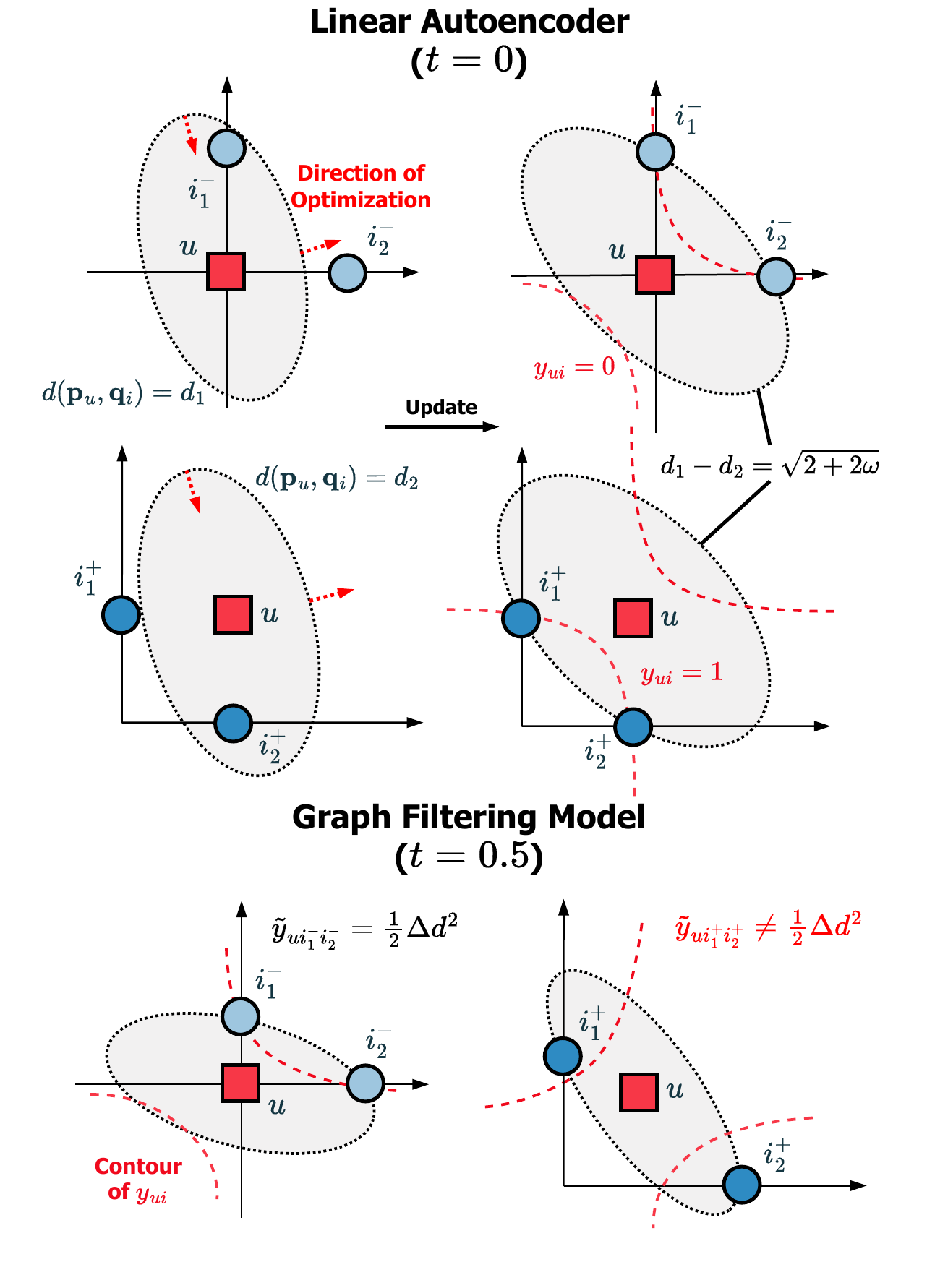}
  \caption{Illustration of signal-based models in the view of metric learning.}
  \label{Fig:Normalization}
  \Description{An illustration of two signal-based models in the view of metric learning. First one is the linear autoencoder, including four subgraphs, corresponding to the case before and after the update of "user-negative item" and "user-positive item", respectively. For "user-negative item", linear autoencoder perform optimization to make all negative item on the same contour. This is also the case for user-positive items, except that the distance between the learned contour of positive items and the user is smaller than that of negative items. Second one is the graph filtering model, the contour of user-item distance is not consistent with the contour of preference score due do the existence of the bias.}
\end{figure}
\subsection{Revisiting Existing Signal-based Models}
Next, we revisit the signal-based models above with considering the learning of generalized Mahalanobis distance metrics. From Eq. \eqref{Eq:LAE} and Eq. \eqref{Eq:GF}, by ensuring the symmetry and zero diagonal of the weight matrix, linear autoencoders and graph filter models can be treated as the case of $t=0$ and $t=0.5$ in our proposed framework, respectively. Figure \ref{Fig:Normalization} visualizes the relations of the preference scores and distances under different cases.

1) Linear autoencoder drives the preference score of interacted item $y_{ui^+}$ to 1 and the preference score of uninteracted item $y_{ui^-}$ to 0 to learn the weight matrix. According to Eq. \eqref{Eq:Diff_GSM}, when $t=0$, this learning goal is equivalent to updating the contour of the distance $d(\mathbf{p}_u, \mathbf{q}_i)$ so that all interacted or uninteracted items are on the same contour. The target contour of the distance for interacted items is smaller than the contour for uninteracted items by $\sqrt{2+2\omega}$. 

2) Graph filtering model shows the case of $t=0.5$, where the above equivalence relationship of distance residual and preference residual does not hold for the interacted items. Since the graph filtering model is training-free, this feature does not affect the producing of recommendations, as the equivalence of $d(\mathbf{p}_u, \mathbf{q}_i)$ and $y_{ui}$ still holds for uninteracted items. However, it poses a problem for the trainable model to apply this normalization strategy, as it will inevitably consider the interacted items.

Thus far, we have revealed the connections between signal-based models and distance metric learning. The different strategies of linear autoencoders and graph filtering models in the normalization strength motivate us to seek to build a trainable signal-based model and generalize it to all cases of $t$. But before that, one question still needs to be answered: What impact does this normalization strength actually have on the results of the recommendation?
\subsection{Effect of Normalization Strength}
\label{SubSec:Effect_Norm}
To identify the effect of normalization strength, we consider the case with involving items with different popularity. When the user $u$ is specified, $y_{ui}$ in Eq. \eqref{Eq:Graph_Signal_Score} can be written as
\begin{equation}
y_{ui}=\mathbf{p}_u^T (d_i^{t}\cdot\mathbf{c}_i),
\end{equation}
where $\mathbf{c}_i$ is the $i$-th column of $\mathbf{C}$. Now, we consider a general optimization problem that introduces non-negative restrictions and $l_2$ regularization, which are commonly adopted in the matrix factorization \cite{Sindhwani2010} and the learning of linear autoencoders \cite{Ning2011, Steck2020a}. The optimization problem of $\mathbf{C}$ is formulated as
\begin{equation}
\underset{\mathbf{C}}{\mathrm{minimize}}{\:\:\mathcal{L}(\mathbf{C})+\lVert \mathbf{C}\rVert_F^2},\:\: s.t.\: \mathbf{C}\geq0,
\end{equation}
where $\mathcal{L}(\mathbf{C})$ is the original loss function for $\mathbf{C}$. The inclusion of $l_2$ regularization results in smaller entries in $\mathbf{C}$, and the non-negative constraint makes all preference score $y_{ui}$ positive. Then, we can easily establish the following connection between preference score and item popularity. When $t$ grows, more popular items with greater $d_i$ are more affected, resulting in larger value of $y_{ui}$. At this point, in order to obtain the same preference score $y_{ui}$, unpopular item requires larger values in $\mathbf{c}_i$, which is being penalized by $l_2$ regularization. Thus, greater $t$ will facilitate signal-based models to generate recommendations of highly popular items. Conversely, a negative $t$ will lead to higher preference scores for less popular items, promoting the novelty of the recommendation. This motivates us to extend the linear autoencoders and graph filtering models to all cases of normalization strength, thus improving the recommended performance while taking novelty into account.

\section{Collaborative Residual Metric Learning}
In this section, we introduce our proposed CoRML model in detail.
\subsection{Triplet Residual Margin Loss}
The triplet margin loss formulated in Eq. \eqref{Eq:Margin_Loss} has been widely used in metric learning models in CF task \cite{Hsieh2017, Park2018, Bao2022}. The objective of triplet margin loss is to keep the distance between dissimilar nodes at least greater than the distance between similar nodes up to the margin $\zeta$. A variation in the margin setting can be seen between different models, which can be a fixed hyperparameter \cite{Hsieh2017} or the trainable parameters for each user and item \cite{Li2020}. As triplet margin loss focuses on the residual of distance, it is consistent with the design of the preference residual. However, according to finding (2) in Section \ref{SubSec:CanLearn}, the always present bias results in an inaccurate reflection of the distance by the preference residual. Meanwhile, the nature of this bias provides us with an idea of using the bias to substitute the margin in the triplet margin loss. Since the bias is always positive and is only proportional to the degree of the interacted items, it can act as an adaptive margin in the loss function. Based on the above discussions, we derive the triplet residual margin loss $\mathcal{L}_{TRM}$ as follows:
\begin{equation}
\begin{aligned}
\mathcal{L}_{TRM}=&\:\sum \limits_{u\in \mathcal{U}}\sum \limits_{i^+\in \mathcal{I}_u, i^-\notin \mathcal{I}_u}(-\tilde{y}_{ui^+i^-})_+\\=&\:\sum \limits_{u\in \mathcal{U}}\sum \limits_{i^+\in \mathcal{I}_u, i^-\notin \mathcal{I}_u}(y_{ui^-}-y_{ui^+})_+,
\end{aligned}
\label{Eq:TRMLoss}
\end{equation}
where $(\cdot)_+$ preserve all positive values and set all negative values to zero. By minimizing the preference residual when the recommendation score of the uninteracted item larger than the interacted item, this loss function will learn a generalized Mahalanobis distance metric to pull the interacted item closer, and push the uninteracted item away and beyond a positive margin. The $\mathcal{L}_{TRM}$ in Eq. \eqref{Eq:TRMLoss} is formulated based on triplets of users and two items. For simplicity, we combine recommendation score terms in different triplets and normalize the weights for each user-item pair. The loss function of $\mathcal{L}_{TRM}$ can be rewritten as
\begin{equation}
\mathcal{L}_{TRM}=\sum \limits_{u\in \mathcal{U}}(\sum \limits_{i^+\in \mathcal{I}_u}\alpha_{ui^+}y_{ui^+}+\sum \limits_{i^-\notin \mathcal{I}_u}\beta_{ui^-}y_{ui^-}).
\label{Eq:UpgradedTRMLoss}
\end{equation}
Here, $\alpha_{ui^+}$ and $\beta_{ui^-}$ are the weights defined as
\begin{equation}
\alpha_{ui^+}=\sum \limits_{i^- \notin \mathcal{I}_u}-\frac{\delta(y_{ui^-}>y_{ui^+})}{\lvert \mathcal{I} \rvert -\lvert \mathcal{I}_u \rvert},\:\:\beta_{ui^-}=\sum \limits_{i^+ \in \mathcal{I}_u}\frac{\delta(y_{ui^-}>y_{ui^+})}{\lvert \mathcal{I}_u \rvert},
\label{Eq:RankingWeight}
\end{equation}
where $\delta(\cdot)$ is the indicator function equals to 1 when the condition is satisfied and 0 otherwise. 
\subsection{Approximated Ranking Weights}
In Eq. \eqref{Eq:UpgradedTRMLoss}, the weights $\alpha$ and $\beta$ are dependent on the ordering relationship of $y_{ui}$ with the same $u$ and different $i$. Since $y_{ui}$ is changed during the optimization, $\alpha$ and $\beta$ need to be updated by sorting $y_{ui}$ of all items at each iteration, incurring highly expensive computational costs. Here, instead of seeking exact numerical values of $\alpha$ and $\beta$, we turn our attention to the relationships between different $(u, i)$ pairs. From Eq. \eqref{Eq:RankingWeight}, given a specific $u$, $\alpha_{ui}$ is always negative and its absolute value decreases with the growth of $y_{ui}$. In contrast, $\beta_{ui}$ is always positive and increases when $y_{ui}$ is growing. This provides us with an idea to approximate $\alpha$ and $\beta$ by the numerical value of $y_{ui}$. Here, we propose the approximated ranking weights $\tilde{\alpha}$ and $\tilde{\beta}$ as
\begin{equation}
\tilde{\alpha}_{ui^+}= \phi y_{ui^+}-1,\: \tilde{\beta}_{ui^-}=\phi y_{ui^-}.
\label{Eq:ApproxWeight}
\end{equation}
Since the original ranking weights $\alpha$ and $\beta$ are normalized to $[-1, 0]$ and $[0,1]$ respectively, we introduce a factor $\phi$ to obtain the similar effect by scaling the preference score $y_{ui}$. The definition of $\phi$ is categorized into the following two components:
\begin{itemize}
    \item Global scaling: Scale all preference scores $y_{ui}$ with a fixed global factor.
    \item User-degree scaling: The user's degree $d_u$ indicates the number of non-zero elements in the user's signal, which may result in the preference scores $y_{ui}$ of different users being in different ranges. For this reason, we use a scaling factor based on the user's degree to adjust the range of $y_{ui}$.
\end{itemize}
Then the scaling factor $\phi$ for user $u$ is then formulated as
\begin{equation}
\phi_u = \epsilon (\frac{d_u}{max_{u \in \mathcal{U}}(d_u)})^{-t_u},
\end{equation}
where $\epsilon$ is the global scaling hyperparameter, and $t_u$ is a normalization factor for user-degree scaling. With a suitable adjustment of $\phi_u$, these approximated weights can then satisfy the conditions discussed above and preserve the relative relationships. By replacing $\alpha$ and $\beta$ in loss Eq. \eqref{Eq:UpgradedTRMLoss} with $\tilde{\alpha}$ and $\tilde{\beta}$ respectively, we obtain the loss function of Collaborative Residual Metric Learning (CoRML) $\mathcal{L}_{CoRML}$ as
\begin{equation}
\mathcal{L}_{CoRML}=\sum \limits_{u\in \mathcal{U}}\sum \limits_{i\in \mathcal{I}}y_{ui}(\phi_u y_{ui}-R_{ui})=tr(\mathbf{Y}^T(\mathbf{\Phi Y}-\mathbf{R})),
\label{Eq:CoRML_Loss}
\end{equation}
where $\mathbf{\Phi}$ is a diagonal matrix containing $\phi_u$ for each user, and $\mathbf{Y} \in \mathbb{R}^{\lvert \mathcal{U} \rvert \times \lvert \mathcal{I} \rvert}$ is the preference score matrix. Then, inspired by the linear autoencoder and graph filtering models, we design a hybrid preference score for CoRML as
\begin{equation}
\mathbf{Y}=\: \mathbf{R} (\lambda \mathbf{D}_I^{-t}\mathbf{H}\mathbf{d}_I^{t}+(1-\lambda)\mathbf{D}_I^{-\frac{1}{2}}\mathbf{G}\mathbf{D}_I^{\frac{1}{2}}),
\label{Eq:CoRML_Score}
\end{equation}
 where $\mathbf{G}=(\mathbf{V}\mathbf{V}^T-diag(\mathbf{V}\mathbf{V}^T))_+$ is obtained by applying positive and diagonal zero constraints to the weight matrix in the graph filtering model. Finally, the optimization problem in CoRML is formulated as 
\begin{equation}
\begin{aligned}
\underset{\mathbf{H}}{\mathrm{minimize}}&{\:\:tr(\mathbf{Y}^T(\mathbf{\Phi Y}-\mathbf{R}))},\\ s.t. &\:\:diag(\mathbf{H})=\mathbf{0}, \mathbf{H}\geq0, \mathbf{H} = \mathbf{H}^T.
\label{Eq:CoRML_Problem}
\end{aligned}
\end{equation}
\subsection{Optimization}
The steps of solving problem in Eq. \eqref{Eq:CoRML_Problem} contain the Sylvester equation, which is not easy to find an closed-form solution. Therefore, we transform the original problem by multiplying both $\mathbf{Y}$ and $\mathbf{R}$ in Eq. \eqref{Eq:CoRML_Problem} by a term $\mathbf{D}_I^{-t}$. The derived problem is shown below, which can be efficiently solved with Alternating Directions Method of Multipliers (ADMM) \cite{Steck2020a, Boyd2011}:
\begin{equation}
\begin{aligned}
\underset{\mathbf{H}, \mathbf{Z}}{\mathrm{minimize}}&{\:\:tr(\mathbf{D}_I^{-t}\mathbf{Y}^T(\mathbf{\Phi Y}-\mathbf{R})\mathbf{D}_I^{-t})+\frac{\theta}{2}  \lVert \mathbf{D}_I^{\frac{1}{2}}\mathbf{H} \rVert^2_F}\\ s.t. &\:\:diag(\mathbf{H})=\mathbf{0}, \mathbf{Z}\geq0, \mathbf{Z} = \mathbf{Z}^T, \mathbf{H}=\mathbf{Z},
\label{Eq:CRML_ADMM}
\end{aligned}
\end{equation}
where $\theta$ is introduced to control the strength of $l_2$ regularization. The regularization term of each row in $\mathbf{H}$ is weighted by $\mathbf{D}_I$ based on their occurrence in $\mathcal{L}_{CoRML}$. Then, $\mathbf{H}$ can be updated by adopting augmented Lagrangian method, and $\mathbf{Z}$ can be updated by the analytic solution of the continuous Lyapunov equation \cite{Bartels1972}. The derived matrix $\mathbf{H}$ will be used to generate the preference score for each user-item pair through Eq. \eqref{Eq:CoRML_Score}.

\section{Experiment}
\subsection{Experimental Setup}
\subsubsection{Datasets and metrics}
We conduct the experiment on four public available datasets: \textit{Pinterest}, \textit{Gowalla}, \textit{Yelp2018} and \textit{ML-20M}. For \textit{ML-20M} dataset, users with at least 5 interactions are retained for consistency with previous studies \cite{Liang2018, Shenbin2020}. The statistics of datasets are summarized in Table \ref{Tab:Dataset}. In each dataset, the interactions are split into train set, valid set and test set with the ratio of 0.6/0.2/0.2.  The model performance are evaluated based on two widely used metrics in CF task: Normalized Discounted Cumulative Gain at $K$ (NDCG@$K$) and Mean Reciprocal Rank at $K$ (MRR@$K$), where $K$ is set to 5, 10 and 20, respectively. 
\begin{table}[!t]
    \centering
    \renewcommand{\arraystretch}{0.6}
    \caption{Statistics of datasets.}
    \label{Tab:Dataset}
    \begin{tabular}{c||cccc}
    \toprule
    Dataset & \#User & \#Item & \#Interaction & Density (\%) \\ \midrule
    Pinterest & 55,187 & 9,916 & 1,463,581 & 0.2675 \\
    Gowalla & 29,858 & 40,981 & 1,027,370 & 0.0840 \\
    Yelp2018 & 31,668 & 38,048 & 1,561,406 & 0.1296 \\
    ML-20M & 136,674 & 13,680 & 9,977,451 & 0.5336 \\ \bottomrule
    \end{tabular}
\end{table}
\subsubsection{Baselines}
Several types of baselines are involved in the performance comparison with CoRML:
\begin{itemize} 
\item  \textbf{Metric learning}: Classical CML \cite{Hsieh2017} and the latest DPCML \cite{Bao2022} designed to promote the diversity of recommendations. In addition, we replace the embeddings in CML with the embeddings produced by graph convolution in LightGCN \cite{He2020} to incorporate graph neighboring information. The model is named L-CML.
\item  \textbf{Autoencoder}: Linear autoencoder SLIM \cite{Ning2011}, EASE \cite{Steck2019}, and non-linear autoencoder RecVAE \cite{Shenbin2020}.
\item  \textbf{Graph filtering model}: GFCF \cite{Shen2021}.
\item  \textbf{GCN model}: UltraGCN \cite{Mao2021} and the state-of-the-art SimGCL \cite{Yu2022} based on contrastive learning.
\end{itemize}
\subsubsection{Hyperparameter Tuning}
To make a fair comparison, we make consistent settings on some key hyperparameters for all comparison models. For all baselines iteratively train the embedding vectors of users and items, optimizer Adam is used with learning rate set to 1e-3, the embedding size is fixed to 64, and the training batch size is set to 4096. For autoencoders and CoRML, the learned weight matrix can be dense or sparse, where the sparsity cannot be explicitly set. To maintain consistency, we perform a sparse approximation \cite{Steck2020a} of the derived matrices $\mathbf{C}$ (equivalent to $\mathbf{H}$ in CoRML) by setting the entries to 0 where $\lvert \mathbf{C}\rvert \leq \gamma$. The threshold $\gamma$ will be adjusted so that the storage size of the sparse matrix $\mathbf{C}_{sparse}$ is less than other types of models with embedding size 64. All sparse matrices are stored in compressed sparse row (CSR) format, which contains approximately twice the parameter numbers as the number of non-zero values (NNZ) in $\mathbf{C}_{sparse}$. For other hyperparameters, a five-fold cross-validation is performed on each model to fine-tune the hyperparameters. For CoRML, $\lambda$ is tuned between 0 and 1 with the step size of 0.05, $\epsilon$ and $\theta$ are tuned in [0.01,0.1,1], $t_u$ is chosen in [0, 0.5, 1], and $t$ is tuned between -0.2 to 0.2 with the step 0.05.

\begin{table*}[!t]
    \centering
    \renewcommand{\arraystretch}{0.6}
    \setul{1pt}{0.4pt}
    \caption{Performance comparison on 4 datasets.}
    \label{Tab:Performance}
    \begin{center}
    \begin{tabular}{cc||cccccccc@{\space\space}c|c}
    \toprule
    Dataset & \multicolumn{1}{c||}{Metric} & CML & L-CML & DPCML & SLIM & EASE & RecVAE & GFCF & UltraGCN & SimGCL & CoRML \\ \midrule
     \multirow{6}{*}{Pinterest} & NDCG@5 & 0.0509 & 0.0594 & 0.0563 & 0.0488 & 0.0558 & 0.0516 & \ul{0.0620} & 0.0572 & 0.0616 & \textbf{*0.0655} \\
     & NDCG@10 & 0.0665 & 0.0766 & 0.0724 & 0.0630 & 0.0704 & 0.0668 & \ul{0.0785} & 0.0729 & 0.0783 & \textbf{*0.0824} \\
     & NDCG@20 & 0.0897 & 0.1021 & 0.0965 & 0.0841 & 0.0921 & 0.0895 & \ul{0.1031} & 0.0962 & \ul{0.1031} & \textbf{*0.1076} \\
     & MRR@5 & 0.1018 & 0.1186 & 0.1133 & 0.0957 & 0.1125 & 0.1024 & \ul{0.1239} & 0.1146 & 0.1237 & \textbf{*0.1306} \\
     & MRR@10 & 0.1164 & 0.1343 & 0.1283 & 0.1084 & 0.1262 & 0.1164 & \ul{0.1390} & 0.1292 & \ul{0.1390} & \textbf{*0.1458} \\
     & MRR@20 & 0.1261 & 0.1444 & 0.1381 & 0.1171 & 0.1353 & 0.1258 & \ul{0.1488} & 0.1387 & \ul{0.1488} & \textbf{*0.1556} \\ \midrule
    \multirow{6}{*}{Gowalla} & NDCG@5 & 0.0853 & 0.0985 & 0.0999 & 0.1100 & 0.1211 & 0.0890 & 0.1174 & 0.1108 & \ul{0.1229} & \textbf{*0.1317} \\
     & NDCG@10 & 0.0953 & 0.1093 & 0.1087 & 0.1156 & 0.1268 & 0.0978 & 0.1257 & 0.1181 & \ul{0.1295} & \textbf{*0.1383} \\
     & NDCG@20 & 0.1125 & 0.1281 & 0.1261 & 0.1302 & 0.1412 & 0.1140 & 0.1440 & 0.1348 & \ul{0.1460} & \textbf{*0.1554} \\
     & MRR@5 & 0.1533 & 0.1743 & 0.1811 & 0.1912 & 0.2186 & 0.1613 & 0.2121 & 0.2001 & \ul{0.2235} & \textbf{*0.2334} \\
     & MRR@10 & 0.1682 & 0.1899 & 0.1957 & 0.2043 & 0.2323 & 0.1752 & 0.2269 & 0.2144 & \ul{0.2377} & \textbf{*0.2479} \\
     & MRR@20 & 0.1768 & 0.1984 & 0.2040 & 0.2118 & 0.2393 & 0.1832 & 0.2352 & 0.2225 & \ul{0.2454} & \textbf{*0.2558} \\ \midrule
    \multirow{6}{*}{Yelp2018} & NDCG@5 & 0.0483 & 0.0574 & 0.0556 & 0.0535 & 0.0611 & 0.0525 & 0.0587 & 0.0585 & \ul{0.0646} & \textbf{*0.0690} \\
     & NDCG@10 & 0.0521 & 0.0617 & 0.0592 & 0.0554 & 0.0628 & 0.0558 & 0.0617 & 0.0621 & \ul{0.0676} & \textbf{*0.0716} \\
     & NDCG@20 & 0.0629 & 0.0742 & 0.0709 & 0.0644 & 0.0722 & 0.0663 & 0.0731 & 0.0737 & \ul{0.0795} & \textbf{*0.0832} \\
     & MRR@5 & 0.1007 & 0.1188 & 0.1156 & 0.1117 & 0.1277 & 0.1106 & 0.1236 & 0.1234 & \ul{0.1349} & \textbf{*0.1435} \\
     & MRR@10 & 0.1149 & 0.1345 & 0.1304 & 0.1245 & 0.1413 & 0.1247 & 0.1380 & 0.1385 & \ul{0.1499} & \textbf{*0.1586} \\
     & MRR@20 & 0.1241 & 0.1443 & 0.1399 & 0.1327 & 0.1496 & 0.1336 & 0.1472 & 0.1478 & \ul{0.1594} & \textbf{*0.1679} \\ \midrule
    \multirow{6}{*}{ML-20M} & NDCG@5 & 0.2319 & 0.2731 & 0.2620 & 0.2785 & 0.3025 & \ul{0.3045} & 0.2718 & 0.2365 & 0.2675 & \textbf{*0.3189} \\
     & NDCG@10 & 0.2326 & 0.2689 & 0.2588 & 0.2710 & 0.2934 & \ul{0.3033} & 0.2671 & 0.2280 & 0.2644 & \textbf{*0.3103} \\
     & NDCG@20 & 0.2486 & 0.2832 & 0.2725 & 0.2813 & 0.3036 & \ul{0.3204} & 0.2799 & 0.2369 & 0.2794 & \textbf{*0.3212} \\
     & MRR@5 & 0.3761 & 0.4341 & 0.4190 & 0.4478 & \ul{0.4829} & 0.4777 & 0.4356 & 0.3919 & 0.4310 & \textbf{*0.4967} \\
     & MRR@10 & 0.3932 & 0.4494 & 0.4347 & 0.4621 & \ul{0.4963} & 0.4923 & 0.4506 & 0.4063 & 0.4466 & \textbf{*0.5098} \\
     & MRR@20 & 0.4002 & 0.4554 & 0.4409 & 0.4677 & \ul{0.5014} & 0.4976 & 0.4566 & 0.4124 & 0.4527 & \textbf{*0.5149} \\ \bottomrule
    \end{tabular}
    \end{center}
    In each metric, the best result is \textbf{bolded} and the runner-up is \ul{underlined}. * indicates the statistical significance of $p<0.01$.
\end{table*}

\subsection{Performance Comparison}
We conduct all experiments with the same Intel(R) Core(TM) i9-10900X CPU @ 3.70GHz machine with a Nvidia RTX A6000 GPU. Table \ref{Tab:Performance} reports the performance comparison on 4 public datasets. The highlights of Table \ref{Tab:Performance} are summarized as follows:

1) Among metric learning baselines, L-CML shows competitive or superior performance compared to the original CML and the latest MF-based metric learning model DPCML. It provides evidence that GCN is effective in capturing higher-order relationships between users and items, as well as bringing performance improvement of metric learning models.

2) Despite the fact that the performance of different baseline methods varies on datasets, trends can be observed based on the types and characteristics of the datasets. On \textit{Gowalla} and \textit{Yelp2018} datasets, GCN models demonstrate better performance among all the baseline models. One possible reason is the balance between the number of user and item. Since GCN models learn embeddings with a fixed length for each user and item, it may experience performance degradation when the number of user and item is unbalanced. 

3) Signal-based models, including graph filtering model GFCF and autoencoders, show superior performance on denser \textit{Pinterest} and \textit{ML-20M} datasets. Different normalization strengths, i.e., the choice of $t$ in signal-based models, may explain the difference of their performance on such two datasets.

4) Overall, our proposed CoRML shows superior performance on all datasets. This can be attributed to the adoption of the idea of metric learning and the extension of the signal-based models. The former achieves similarity propagation by learning a valid distance metric, and the latter can capture various characteristics of signals under different normalization strength.

\subsection{Benefits of CoRML}
\subsubsection{Mitigating item popularity bias}
\begin{figure}[t!]
  \centering
  \includegraphics[width=1.65in]{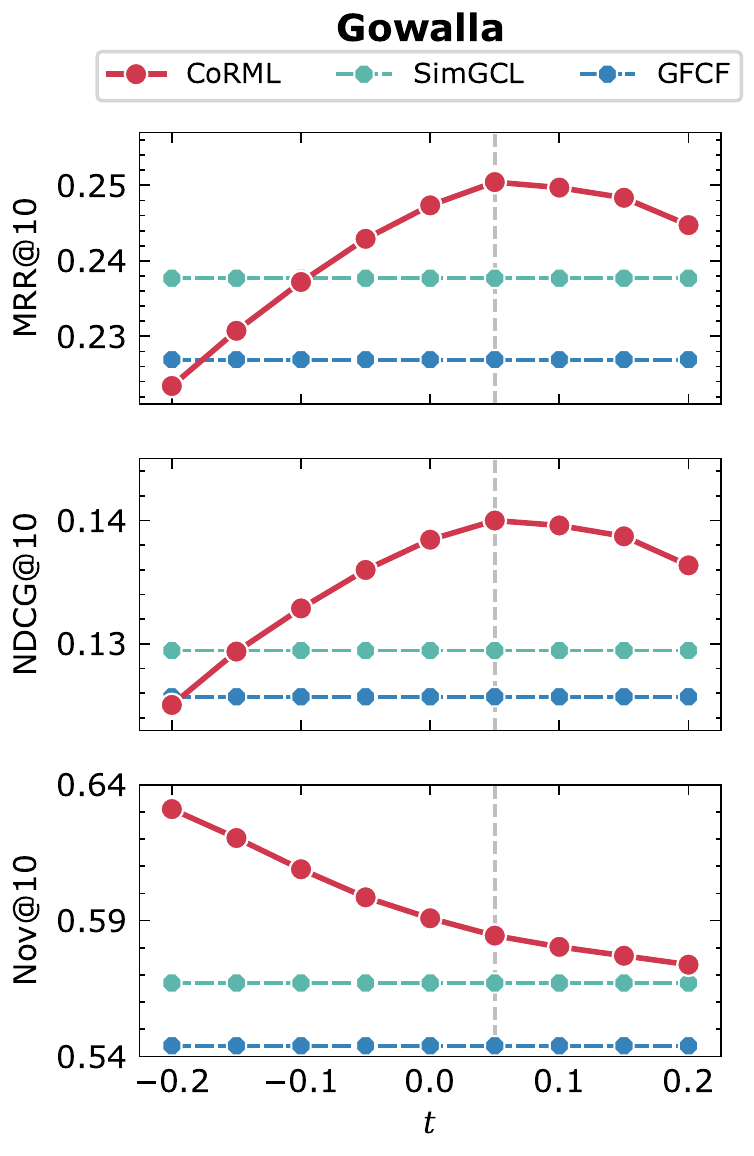}
  \hfil
  \includegraphics[width=1.65in]{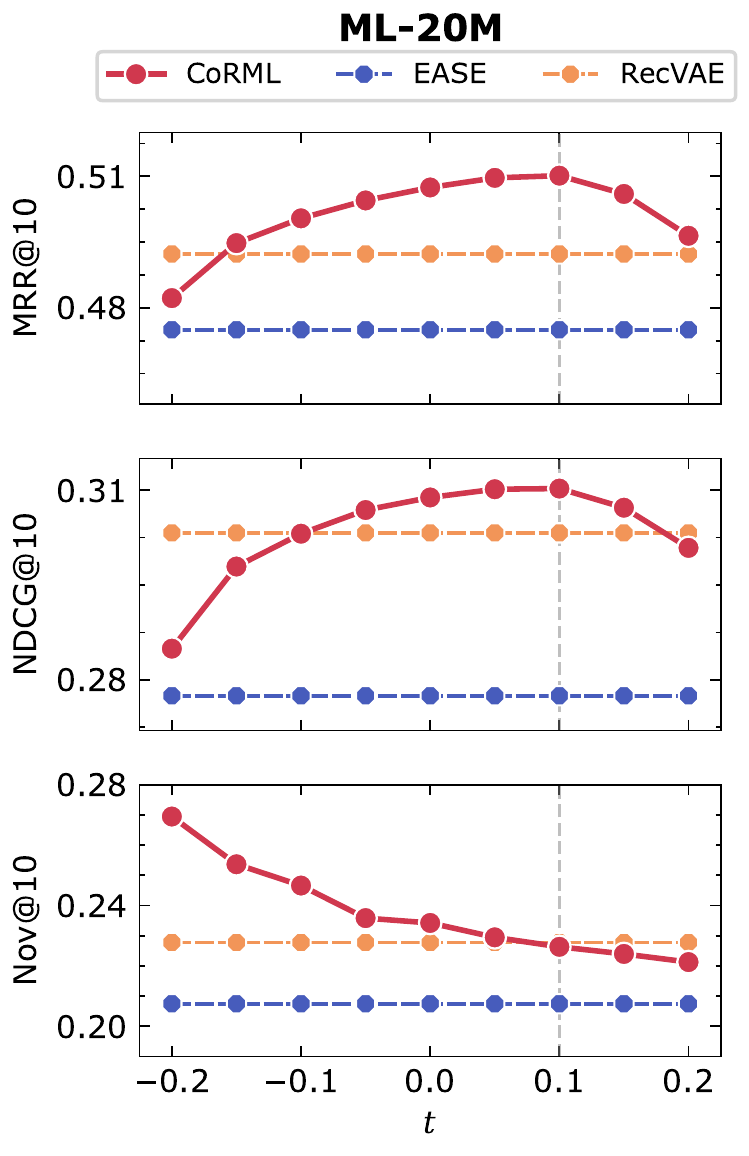}
  \caption{Effects of the normalization strength $t$ on CoRML.}
  \Description{Line graph showing the NDCG@10, MRR@10, and Nov@10 against of the hyperparameter t from -0.2 to 0.2 in increments of 0.05 on the X axis. Six subgraphs are shown, corresponding to 2 test datasets: Gowalla and ML-20M. Each dataset contains three subgraphs, corresponding to 3 metrics. In subgraphs of MRR@10 and NDCG@10, the performance of CoMRL first rise then fall when t is increased. The best performance of CoRML are higher than the compared baselines: SimGCL and GFCF on Gowalla dataset, and EASE and RecVAE on ML-20M dataset. For Nov@10, CoRML shows a trend of decline on both datasets when t grows. CoRML achieves higher or equal Nov@10 compared to baseline models when MRR@10 and NDCG@10 are reach maximum.}
  \label{Fig:Novelty}
\end{figure}

In recent works in CF, the popularity bias has been brought to light in recommendation scenarios \cite{Zhu2021, Zhao2022}. In CoRML, the normalization strength $t$ has been previously discussed to be associated with the popularity of items in Section \ref{SubSec:Effect_Norm}. In order to ascertain how $t$ affects the performance and novelty of recommendations, we conduct experiments on CoRML and two chasing baselines. The performance of the recommendation is still measured by MRR and NDCG, while the novelty is measured by a metric introduced by \cite{Zhou2010} as:
\begin{equation}
Nov@K=\frac{1}{\lvert \mathcal{U} \rvert K}\sum \limits_{u\in \mathcal{U}}\sum \limits_{i\in \mathcal{I}_u'(K)}-\frac{1}{log_2 \lvert \mathcal{U} \rvert} log_2\frac{d_i}{\lvert \mathcal{U} \rvert},
\end{equation}
where $\mathcal{I}_u'(K)$ is the top-$K$ items recommended for user $u$. Figure \ref{Fig:Novelty} shows the results on the \textit{Gowalla} and \textit{ML-20M} datasets.
\begin{itemize}
\item Clearly, the novelty of the recommendations of CoRML is decreasing as $t$ increases from -0.2 to 0.2, indicating more popular items are recommended to users. These results provide further support for the discussion in Section \ref{SubSec:Effect_Norm}, indicating the ability of CoRML to control the novelty of recommendation and reduce popularity bias.
\item In contrast to the monotonic variation of novelty with $t$, there is a peak in the performance of the recommendations when $t$ varies between -0.2 and 0.2. For MRR and NDCG metrics, CoRML showed differences in two tested datasets, with optimal values obtained at $t=$0.05 and $t=$0.1 respectively. It shows that the performance and novelty of the recommendations are not just trade-offs. 
\item When compared to baselines with item popularity taken into account, CoRML can ensure superiority both in performance and novelty. This phenomenon suggests that it is possible to ensure recommendation performance while considering novelty in recommendation scenarios of different natures by the fine-tuning of $t$. 
\end{itemize}

\subsubsection{Efficient training}
\begin{figure}[t!]
  \centering
  \includegraphics[width=2.0in]{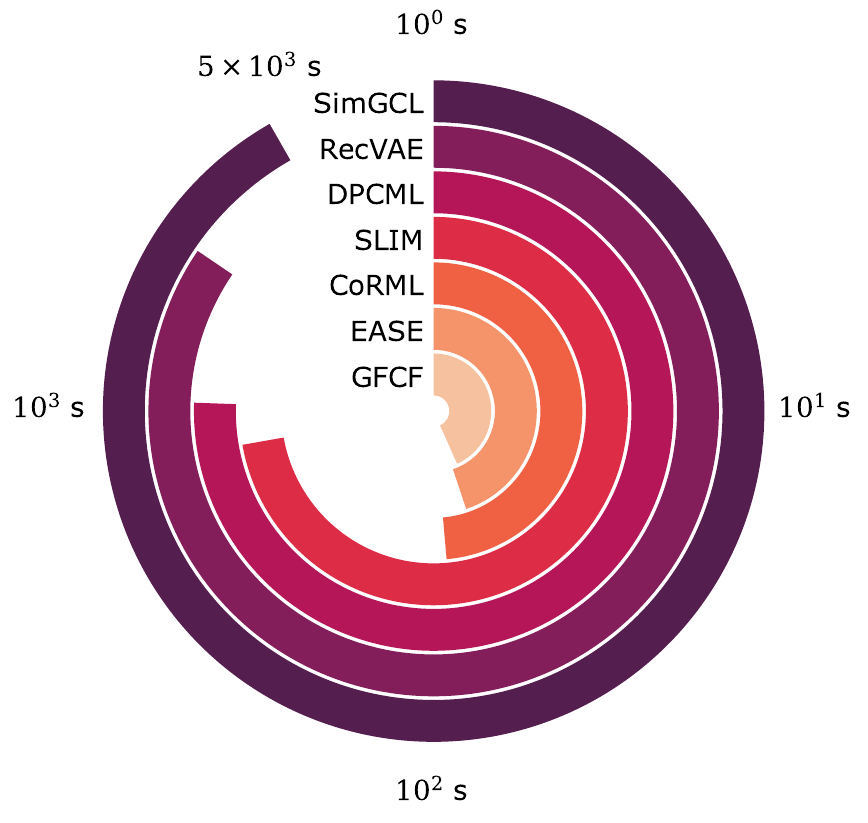}
  \caption{Comparison of training time on \textit{ML-20M} dataset.}
  \Description{A bar plot showing the training time of different models on ML-20M dataset. The longest training times are for SimGCL and RecVAE with over 1000 seconds. They are closely followed by DPCML and SLIM, reaching hundreds of seconds. Finally, CoRML, EASE and GFCF take only a few tens of seconds to train.}
  \label{Fig:Efficiency}
  \end{figure}

Figure \ref{Fig:Efficiency} shows the training times of representative baselines on the large-scale dataset \textit{ML-20M}. It can be observed that the latest GCN model SimGCL, the non-linear autoencoder RecVAE, and the MF-based metric learning model DPCML take a substantial amount of time in training. The reason for this is that they require multiple iterations of optimization in mini-batch. For comparison, signal-based models such as EASE and GFCF achieve up to tens of times higher efficiency. At the same time, our proposed CoRML is based on an extension of signal-based models and eventually formulates an optimization problem with a similar form. Consequently, it also achieves high efficiency in the training process, with training times in the same order of magnitude as the most efficient baselines. Compared to existing CML models, CoRML retains the effective part of a distance metric through residual learning, which retains the advantages of metric learning and brings significant efficiency gains.

\subsection{Detailed Study of CoRML}
\subsubsection{How do weights of preference score affect the performance?}
\begin{figure}[t!]
  \centering
  \includegraphics[width=1.65in]{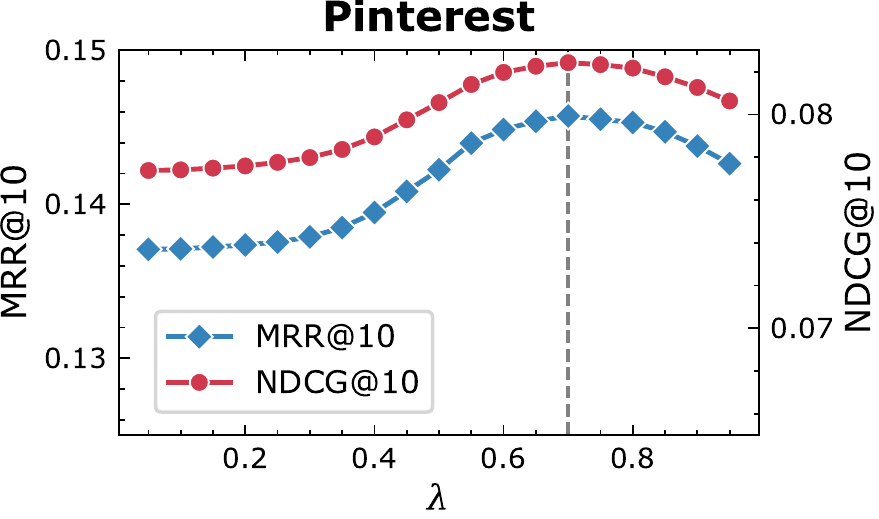}
  \hfil
  \includegraphics[width=1.65in]{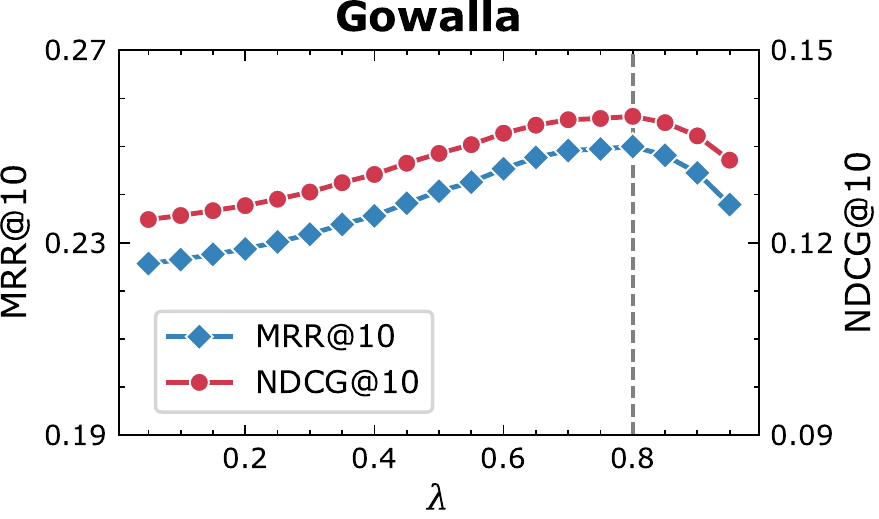}
  \hfil
  \includegraphics[width=1.65in]{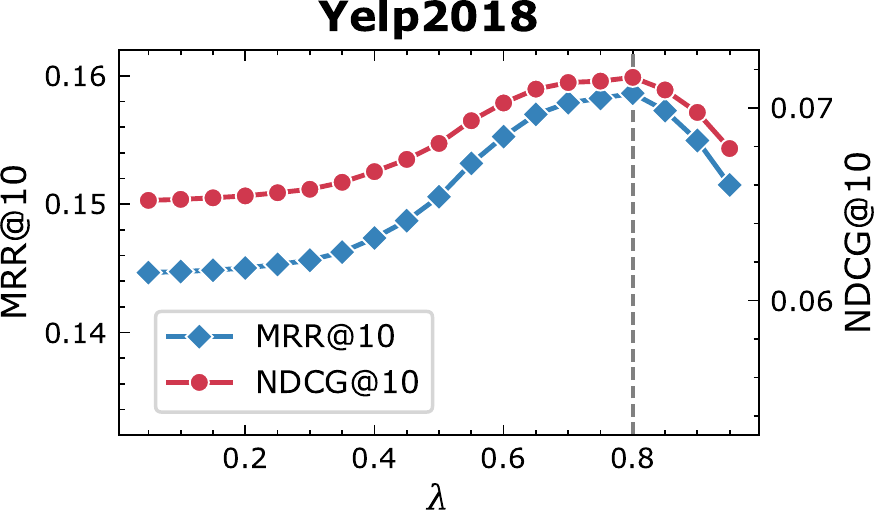}
  \hfil
  \includegraphics[width=1.65in]{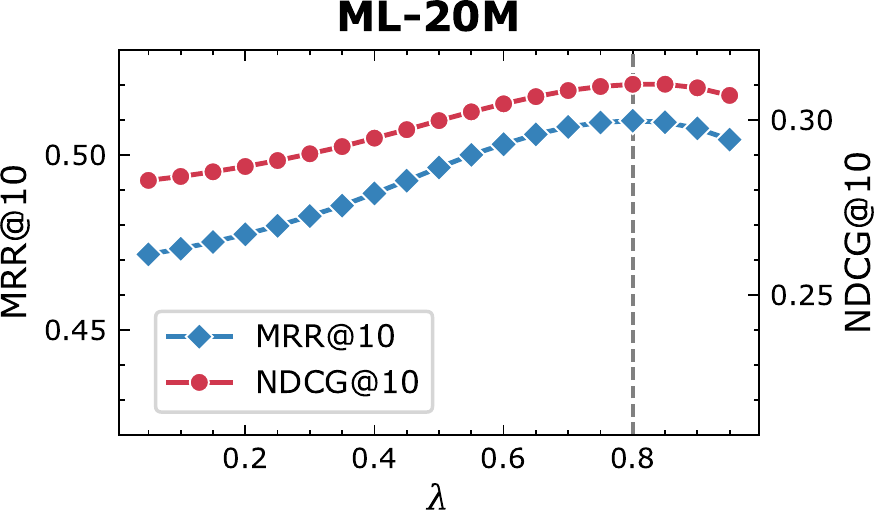}
  \caption{Effect of weighting factor $\lambda$ on CoRML.}
  \Description{Line graph showing the MRR@10 and NDCG@10 against of the hyperparameter lambda from 0.1 to 0.9 in increments of 0.05 on the X axis. Four subgraphs are shown, corresponding to four test datasets: Pinterest, Gowalla, Yelp2018, and ML-20M. In all subgraphs, both two lines for Recall@20 and NDCG@20 increase first and then decrease when lambda grows from 0.1 to 0.9. Recall@20 and NDCG@20 peaks in different subgraphs with different value of lambda ranging from 0.4 to 0.8.}
  \label{Fig:Lambda}
  \end{figure}

As shown in Eq. \eqref{Eq:CoRML_Score}, the preference score in CoRML is formulated based on $\mathbf{H}$ and $\mathbf{G}$, which are balanced by hyperparameter $\lambda$. To investigate the effect of weighting residuals on the model performance, we conduct experiments on all test datasets. Figure \ref{Fig:Lambda} shows the variation of performance when $\lambda$ is tuned between 0.05-0.95. It can be observed that when $\lambda <$0.5, performance is consistently poor across all datasets. Increasing $\lambda$ results in optimum performance when $0.6 \leq \lambda \leq 0.8$, followed varying degrees of drop. \textit{ML-20M}, the dataset with the highest density,  exhibits the slightest drop. This possibly provides an evidence for the inference that the graph filtering residual focused on the low-rank approximation of weight matrix is more important on sparse dataset, where users and items have fewer interactions. 

\subsubsection{Effect of approximated ranking weights}
\begin{figure*}[t!]
  \centering
  \includegraphics[width=\linewidth]{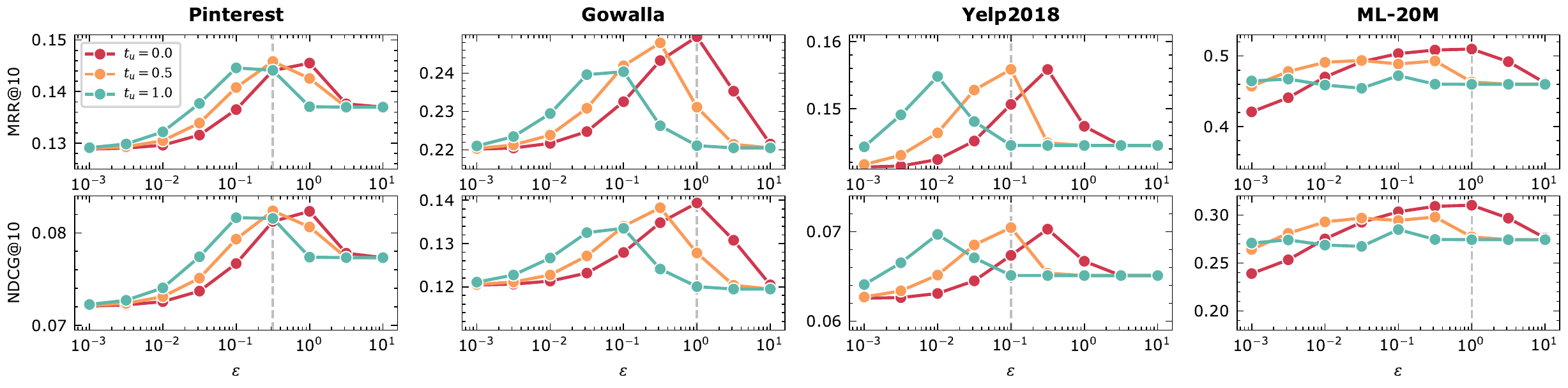}
  \caption{Effect of the approximated ranking weights $\phi$ on CoRML.}
  \Description{Line graph showing the MRR@10 and NDCG@10 against of the hyperparameter epsilon from 0.001 to 10 in the logarithm scale on the X axis. Eight subgraphs are shown, corresponding to four test datasets: Pinterest, Gowalla, Yelp2018, and ML-20M. Each subgraph includes 3 lines, representing different chosen values of hyperparameter "t subscript u": 0, 0.5, and 1. In all subgraphs, all lines increase first and then decrease when epsilon grows from 0.001 to 10. MRR@10 and NDCG@10 peaks between 0.1 to 1. Set "t subscript u" to 0.5 achieves an overall all best performance on all three choices.}
  \label{Fig:Epsilon}
  \end{figure*}
In CoRML, approximated ranking weights $\tilde{\alpha}$ and $\tilde{\beta}$ formulated as Eq. \eqref{Eq:ApproxWeight} are designed to facilitate training without ranking all items. The weights $\tilde{\alpha}$ and $\tilde{\beta}$ are scaled by factor $\phi_u$ to act on the optimization of positive and negative user-item pairs, respectively. The factor $\phi_u$ is controlled by $\epsilon$ as global scaling hyperparameter and $t_u$ as user-degree scaling factor, whose effects are tested and shown in Figure \ref{Fig:Epsilon}. We can make the following observations: 1) The role of $t_u$ is to enlarge preference scores in varying user degrees. Therefore, increasing $t_u$ in general makes the value of $\epsilon$ smaller when the performance is optimal. 2) Varying $t_u$ has a small effect on the optimal performance of CoRML. Setting $t_u = 0.5$ can result in excellent performance on all datasets. 3) Global scaling factor $\epsilon$ shows more significant impact on the model performance. On the logarithmic scale, extremely large or extremely small $\epsilon$ may deteriorate the model performance. This can be justified through the definition of the approximated ranking weights. As shown in Eq. \eqref{Eq:ApproxWeight}, a very small $\epsilon$ makes all positive user-item pairs have nearly the same weight -1, while a very large $\epsilon$ can reverse the sign of $\tilde{\alpha}$, deviating the learning objective. In general, keeping $\epsilon$ in the range of $[0.1,1]$ can lead to good recommendation performance.
  
\section{Related Works}
\subsection{Collaborative Filtering}
Collaborative Filtering (CF) \cite{Adomavicius2005} has become a popular research topic in the Internet era. Since CF can be considered as a task to complete entries in the user-item interaction matrix, Matrix Factorization (MF) \cite{Koren2009}, as a strategy for matrix completion, naturally becomes the foundation of the mainstream approach in CF. MF assumes that the user-item interaction matrix is low-rank and can be recovered by learning the embedding vectors of users and items. Most MF models generate predicted entries by the dot product of user and item embedding vectors, while they can be optimized by minimizing the error of individual entries \cite{Koren2009} or maximizing the difference between positive and negative samples \cite{Rendle2009}. They are both widely adopted in the subsequent proposed methods which introduce refined structures like neural networks \cite{Xue2017, He2018}. These approaches enable a light design by focusing on the modeling of single user-item entry, while neglecting the synergy between different interactions. In recent years, this type of global relationships has gradually received more attention and has been incorporated in CF models in the form of interaction graph \cite{Zheng2018}. With the emerging of Graph Convolutional Networks (GCN), GCN models \cite{Wang2019, He2020} quickly become the state-of-the-art in MF-based models and are continuously improved to achieve advances in efficiency and accuracy \cite{Mao2021, Yu2022}.

Unlike MF-based models, another class of method implements CF by treating the user's historical interactions as features, and modeling the relationships between items \cite{Deshpande2004, Wei2023a}.  A classical approach is the linear autoencoder \cite{Ning2011, Steck2019, Steck2020a}, which models an item-item relationship matrix to encode user features. This idea is then extended by subsequent studies and applied to nonlinear denoising autoencoders \cite{Wu2016} and variational autoencoders \cite{Liang2018, Shenbin2020}. A recent work \cite{Shen2021} considers CF in terms of graph signal processing and proposes a framework for signal-based models, which can incorporate the linear autoencoders and the ideal case of MF and GCN models. They also propose a simple but effective graph filtering model GFCF to model item relationships. 

\subsection{Metric Learning}
Metric learning \cite{Xing2002, Ghodsi2007} learns a distance metric to fit the distance and similarity between training data: separating dissimilar samples and pushing similar samples closer. Over the decades, metric learning has gained attention and adoption in many fields, such as Computer Vision \cite{Weinberger2009} and Nature Language Processing \cite{Lebanon2006}. In CF, the recent progress related to metric learning has been mainly influenced by work \cite{Hsieh2017}. In \cite{Hsieh2017}, based on the idea of MF, the authors propose a metric learning framework CML to estimate the user-item relationship by the distance of embedding vectors in Euclidean space. CML is then adopted and improved in subsequent studies by incorporating translation vectors \cite{Park2018}, adopting adaptive margin \cite{Li2020}, and promoting diversity \cite{Bao2022}. 

As discussed, existing studies of metric learning on CF have been conducted over the distance metric on Euclidean space. In a recent survey of metric learning \cite{Ghojogh2022}, the authors formulate a typical metric learning problem as the learning of generalized Mahalanobis distance, and show that the Euclidean distance is a special case of generalized Mahalanobis distance. On the other hand, CML and its subsequent studies are based on MF and do not involve signal-based models. This provides us with motivation and becomes the major difference between our work and existing works.

\section{Conclusion}
In this paper, we delve into the signal-based model, unveil its connection to the distance metric, and finally propose a novel CoRML model. In particular, we identify the preference scores in signal-based models are strongly tied to the residuals of distance between user and different items. We also found that the normalization strengths of user interaction signals have an explicit effect on the novelty of recommendation, which is neglected by existing works. By leveraging connections between preference scores and distance residuals, CoRML is able to capture fine-grained user preferences with full advantages of metric learning. Moreover, it yields high training efficiency through introducing a novel approximated ranking weight. A comprehensive comparison with existing CF models shows advantages of CoRML in terms of performance, efficiency and novelty, validating the role of metric learning in signal-based CF models.
\begin{acks}
This work was partially supported by the Research Grants Council of the Hong Kong Special Administrative Region, China (Project No. CityU 11216620), and the National Natural Science Foundation of China (Project No. 62202122).
\end{acks}

\bibliographystyle{ACM-Reference-Format}
\balance
\bibliography{CoRML}

\appendix

\end{document}